\documentclass[%
 reprint,
unsortedaddress,
 amsmath,amssymb,
 aps,
]{revtex4-2}

\usepackage{hyperref}

\usepackage{pdfpages} 

\makeatletter
\AtBeginDocument{\let\LS@rot\@undefined}
\makeatother

\usepackage{graphicx}
\usepackage{dcolumn}
\usepackage{bm}
\newcommand{\n}{\nonumber}

\usepackage[capitalize]{cleveref}

\begin{document}

\title{Gravity-controlled non-equilibrium Casimir pressure in a binary liquid mixture}

\author{Marcin~Piotr Pruszczyk}
\email{mpruszcz@sissa.it}
\affiliation{%
 SISSA --- International School for Advanced Studies, via Bonomea 265, 34136 Trieste, Italy}%
\affiliation{%
 INFN, via Bonomea 265, 34136 Trieste, Italy}%

\author{Roberto Cerbino}
\affiliation{
Faculty of Physics, University of
Vienna, Boltzmanngasse 5, 1090 Vienna, Austria}%

\author{Andrea Gambassi}
\affiliation{%
 SISSA --- International School for Advanced Studies, via Bonomea 265, 34136 Trieste, Italy}%
\affiliation{%
 INFN, via Bonomea 265, 34136 Trieste, Italy}%

\date{\today}

\begin{abstract}

We investigate the non-equilibrium Casimir pressure in an isothermal binary liquid mixture maintained in a spatially constant and stationary concentration gradient parallel to gravity and confined within a three-dimensional slab of thickness $L$, bounded by two infinite plates parallel to both the gravitational field and the imposed gradient. 
We assume that the liquid mixture, under the same non-equilibrium conditions, occupies both the interior and the exterior of the slab.
Using fluctuating hydrodynamics, we show that the resulting finite-size excess pressure on the plates is described by a scaling function of the dimensionless variable
$k_{\mathrm{RO}}L$, where $k_{\mathrm{RO}}$ is the gravity-induced roll-off wavevector. 
At large separations, 
this Casimir pressure decays as $1/(k_{\mathrm{RO}}L)$. Depending on the thermodynamic properties of the mixture, the corresponding force can be either attractive or repulsive, while it vanishes for ideal solutions. Since the mixture is assumed to be far  
from its consolute critical point, 
the Casimir pressure investigated here is entirely of non-equilibrium origin and it vanishes in the absence of the imposed concentration gradient. Finally, we propose an experimental 
setup where this force might be measured,
consisting of 
two optically trapped colloidal particles immersed in a dense aqueous colloidal suspension diffusing into an overlying layer of pure water, estimating the expected magnitude of the resulting force.
\end{abstract}
\maketitle


\section{Introduction}

Fluctuation-induced forces arise from the confinement of fluctuations with long-range correlations by bodies imposing boundary conditions (BCs). Originally predicted in quantum electrodynamics, where they result from the confinement of vacuum fluctuations of the electromagnetic field \cite{Casimir_1948}, analogous forces also emerge in soft matter, most notably near second-order phase transitions \cite{Fisher_1978, gambassi_critical_2024}.
In this context, the forces stem from confining the \emph{equilibrium} thermal fluctuations of the order parameter of such critical point,  which are characterized by long-range correlations.
A binary liquid mixture close to its demixing point provides a concrete example \cite{Gambassi_2009_exp}.
Crucially, these so-called critical Casimir forces \cite{Gambassi_2009} were  experimentally measured, first indirectly in the film geometry for quantum \cite{Garcia_1999} and classical \cite{Fukuto_2005} liquids, 
and then directly,
acting on a colloidal sphere close to a flat surface \cite{Hertlein_2008}.
{Beyond the case of critical points at equilibrium, fluctuations with long-range correlations emerge generically in statistical systems out of equilibrium \cite{Spohn_1983, Derrida_2007}.}
Consequently, similar fluctuation-induced forces are expected to act on 
bodies which impose BCs at their surfaces on these fluctuations.  This was shown to be the case, among others, in driven diffusive systems with conserved dynamics \cite{Aminov_2015}, in systems after a temperature  or activity quench 
\cite{Rohwer_2017_PRL, Rohwer_2017}
in driven electrolytes \cite{Mahdisoltani_2021_PRL, Mahdisoltani_2021_NJP}, 
in the presence of a spatially inhomogeneous temperature \cite{Najafi_2004}, or in systems intrinsically out of equilibrium \cite{Caprini_2018}, e.g.,
in active flocking matter \cite{Fava_2024}. 

In non-equilibrium (NE) steady states with stationary gradients of temperature or concentration, the  fluctuations arising from the coupling between 
hydrodynamic modes are 
naturally described within the framework of 
fluctuating hydrodynamics \cite{Landau_1959_hydro, Ortiz_de_Zarate_book}. 
Within this approach, a NE contribution to the bulk pressure in confined liquids and liquid mixtures was derived in Refs.~\cite{Kirkpatrick_Giant_2013, Kirkpatrick_Ortiz_de_zarate_2014, 
Kirkpatrick_Ortiz_de_zarate_2015_Cas, kirkpatrick_ortiz_de_Zarate_2016, ortiz_de_zarate_non-equilibrium_2015, Kirkpatrick_Ortiz_de_zarate_2016_Cas}. 
There, the gradient was assumed to be perpendicular  to the confining plates constituting a slab of thickness $L$, and filling solely the space inside the slab. 
This  
additional NE pressure was thus the excess pressure with respect to the system in equilibrium and, 
in the absence of gravity, it turned out to depend linearly on $L$.

\begin{figure}
\centering
\includegraphics[width=.9
\linewidth]{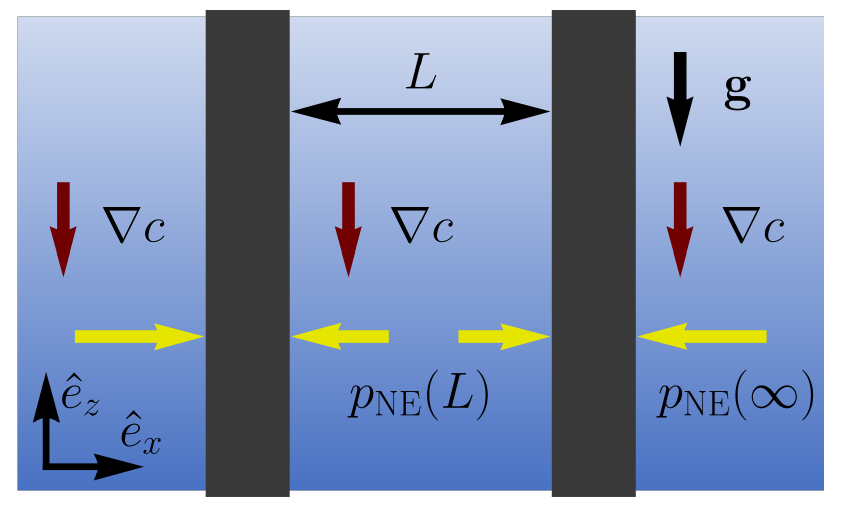}
\label{fig:wide}
\caption{
Cartoon of the physical setup considered in this work. The gradient $ \mathbf{\hat{e}}_z\nabla c$ (with $\nabla c<0$) of the concentration $c$ of the denser solute in the less dense solvent,  represented by the blue color gradient, is parallel to the gravitational acceleration $-\mathbf{\hat{e}}_zg$ and to two parallel plates separated by a distance $L$. The mixture in the presence of the gradient fills the space also outside the plates. 
The surface of the plate which is in contact with the liquid within the slab is subject to the non-equilibrium pressure $p_{\rm NE}(L)$, while the other surface is subject to the pressure $p_{\rm NE}(L \to \infty)$ (yellow arrows). Accordingly, the net pressure acting on each plate is $p_{\rm Cas}(L) = p_{\rm NE}(L) - p_{\rm NE}(L \to \infty)$.
}  
\label{fig:setup}
\end{figure}

In this work, we derive 
the NE Casimir pressure $p_{\mathrm{Cas}}(L)$ generated by thermal
hydrodynamic fluctuations 
in an isothermal binary liquid mixture confined within a three-dimensional slab of thickness $L$,  
occurring in the presence of both a non-equilibrium concentration gradient and gravity.
In particular, differently from Refs.~\cite{Kirkpatrick_Giant_2013, Kirkpatrick_Ortiz_de_zarate_2014, Kirkpatrick_Ortiz_de_zarate_2015_Cas, kirkpatrick_ortiz_de_Zarate_2016, ortiz_de_zarate_non-equilibrium_2015, Kirkpatrick_Ortiz_de_zarate_2016_Cas},
mentioned above, we consider the case in which both the gradient $\bm{\nabla} c = \mathbf{\hat{e}}_z\nabla c $ of the concentration $c$ of the solute in the solvent,
and the gravitational acceleration $\mathbf{g} = -\mathbf{\hat{e}}_zg$ are parallel to the confining walls, and the liquid mixture is present under the same physical conditions both inside and outside the slab, as sketched in Fig.~\ref{fig:setup}.
%
This setting (similar to that considered in Ref.~\cite{Aminov_2015})  allows us to define the NE Casimir pressure $p_{\mathrm{Cas}}(L)$ as the excess contribution, relative to the bulk $L\to\infty$, to the finite-size NE pressure $p_{\rm NE}(L)$. 
To guarantee that the system is far from  convective instability, we assume that the solute is denser than the solvent and that its concentration is larger than the equilibrium value at the bottom of the slab while being smaller at the top, i.e., that $\nabla c< \nabla c_{\rm gr}<0$, where $\nabla c_{\rm gr}$ is the equilibrium gravity-induced gradient determined by, c.f., Eq.~\eqref{c_grav}. 

We show that gravity {determines} the rate of decay of the NE Casimir pressure $p_{\mathrm{Cas}}(L)$ with increasing system size $L$ 
via a 
dimensionless scaling variable $k_{\rm RO}L$, where $k_{\rm RO}$ is the so-called roll-off wavevector (see, e.g., Ref.~\cite{giavazzi_soft_matter_2016}), i.e.,
\begin{equation}
    p_{\mathrm{Cas}}(L) =
\frac{1}{2}\mathcal{A}\,S_E\left(\frac{\nabla c}{\nabla c_{\mathrm{gr}}} -1 \right)k^3_{\mathrm{RO}} \Psi^{\rm(fs)}(k_{\rm RO} L).
    \label{eq:result_anticipate}
\end{equation}
Here $\Psi^{\rm(fs)}(X)$ is a finite-size scaling function, while $\mathcal{A}\,S_E$ is a thermodynamic prefactor discussed below, c.f., Eqs.~\eqref{p_NE} and \eqref{eq:p_NE_result}, the sign of which depends on the material properties of the mixture \cite{Kirkpatrick_Ortiz_de_zarate_2016_Cas}. Note that this prefactor vanishes for ideal solutions, as we discuss in the Supplemental Material (SM),
while it takes a particularly simple form for aqueous colloidal suspensions, see, c.f., Eqs.~\eqref{eq:p_NE_result_colloid} and \eqref{eq:s_def}.
As anticipated, 
$p_{\mathrm{Cas}}(L)$ in Eq.~\eqref{eq:result_anticipate} vanishes at equilibrium, i.e., for $\nabla c = \nabla c_{\rm gr}$
Moreover, at large separations $L$ between the plates, the pressure in three spatial dimensions turns out to decay algebraically $\sim L^{-1}$.
Finally, we note that the 
roll-off wave vector $k_{\rm RO}$ has been measured experimentally \cite{giavazzi_soft_matter_2016} via a light-scattering experiment in a dense aqueous colloidal suspension,  i.e., LUDOX TMA silica particles dispersed in water, diffusing into an overlying layer of pure water. 
As we discuss further below, in such a binary mixture, 
the fluctuation-induced pressure acting on two trapped colloidal particles separated by a distance of the order of the inverse roll-off wavevector are, in principle, within experimental reach.

\section{The model}
\label{model}

{In what follows, we consider the velocity field $\mathbf{v}$, the mass fraction $c  = \rho_1/\rho$ of the solute, where $\rho_1$ is the mass density of the solute and $\rho_2$ is the mass density of the solvent, and the  total mass density  $\rho =\rho_1 +\rho_2$ of the mixture as dynamic variables of the model.} 
For future convenience, we note that the thermodynamic variable conjugate to the mass fraction $c$ is the relative specific (i.e., per unit mass) chemical potential $\mu$, i.e., the difference $\mu = \mu_1 -\mu_2$ between the specific  chemical potentials $\mu_1$ and $\mu_2$ of the solute and solvent, respectively.
While the model introduced below holds for any binary liquid mixture, the physical system  we have in mind is that of aqueous colloidal suspensions, in which tiny colloidal particles (e.g., silica) are dispersed in pure water. In this case, $c$ denotes the mass fraction of the tiny colloids in the mixture, and $\delta c$ its local (thermal) fluctuation from the average value.

We assume below 
that the fluctuations of the density of the mixture are so small that they do not significantly affect the inertia or mass conservation of the flow, but only the gravitational force which drives buoyancy.
Within this (Boussinesq) approximation, 
the fluctuating hydrodynamics equations for  the velocity fluctuations $\delta \mathbf{v}(\mathbf{r}, t)$ and the concentration fluctuations  $\delta c(\mathbf{r}, t)$ in an isothermal binary mixture are \cite{Ortiz_de_Zarate_book, Segre_1993} 
\begin{align}
\bm{\nabla}\cdot\mathbf{\delta v}(\mathbf{r}, t) &= 0, \label{eq:NS_incomp} \\
    \left( \partial_t - \nu \nabla^2 \right) \delta\mathbf{ v}(\mathbf{r}, t)   -\beta \mathbf{g} \delta c(\mathbf{r}, t) &= \frac{1}{\rho}\bm{\nabla}\cdot\mathbf{\delta\Pi}(\mathbf{r}, t), \label{eq:NS_moment}  
    \\  \left( \partial_t - D \nabla^2 \right)\delta c(\mathbf{r}, t) + \mathbf{\delta v}(\mathbf{r}, t)\cdot\bm{\nabla} c   &= -\frac{1}{\rho}\bm{\nabla}\cdot\mathbf{\delta J}(\mathbf{r}, t).\label{eq:NS_diffusion}
\end{align}

These equations also neglect the effects of thermal expansion and temperature fluctuations and assume that the average velocity field vanishes, i.e., $\mathbf{v} = 0$, such that only its fluctuations $\delta\mathbf{v}$ matter.
{In Eq.~\eqref{eq:NS_moment},} $\nu$ is the kinematic viscosity, 
\begin{equation}
    \beta = \rho^{-1}\left( \partial \rho /\partial  c\right)_{p,T}
\label{eq:def-b}
\end{equation} 
is the solutal expansion coefficient, and $\mathbf{g} = -\mathbf{\hat{e}}_zg$ is the gravitational acceleration. {In Eq.~\eqref{eq:NS_diffusion}, instead, $D$ is the diffusion coefficient of the solute in the solvent}, while $\bm{\nabla}c$ is the deterministic, externally imposed NE concentration gradient (which we assume to be pointing along the $z$-direction).  
This approximately constant concentration gradient is realized, e.g., in free-diffusion experiments in which a dense colloidal suspension is overlaid with pure solvent \cite{giavazzi_soft_matter_2016}, or, at least in principle, by coupling the system to two reservoirs at different concentrations, as considered in Ref.~\cite{Aminov_2015}. In the former case, the gradient evolves on the diffusion timescale, which is much slower than the relaxation of the fluctuations, rendering the 
quasi-stationary approximation appropriate.

In the presence of gravity, the mixture equilibrates to a state characterized by the constant gravity-induced concentration gradient
\begin{equation}
\bm{\nabla} c_{\mathrm{gr}} =\mathbf{\hat{e}_z} {\nabla} c_{\mathrm{gr}} =-\mathbf{\hat{e}_z} \beta g\chi,
\label{c_grav}
\end{equation}
where $\chi  = \left(\partial c / \partial \mu \right)_{p,T}$ is the osmotic susceptibility and $g = |\mathbf{g}|$.
This equality follows the fact that, at mechanical equilibrium, the mean concentration 
$c$ is determined by the balance between the osmotic pressure gradient $\nabla c_{\mathrm{gr}} / \chi$ and the buoyancy $-\beta g$.
In the case of aqueous colloidal suspensions, the tiny particles are denser than water (i.e., $\beta >0$, see \cref{eq:def-b}) and, at equilibrium, their concentration is larger at the bottom of the system. Accordingly, when $\nabla c> 0$ (note $\nabla c_{\rm gr}<0$), the mixture above is denser than the mixture below, which leads to convective instability, when gravitational effects (buoyancy) dominate diffusion  in the dynamics of $c$. Moreover, in the experiments of Ref.~\cite{giavazzi_soft_matter_2016}, the silica particles were diffusing into an overlying layer of pure water. {Accordingly,  to make a connection to that experiment} and to avoid convective instabilities, we assume  $\nabla c < \nabla c_{\rm gr}$ in what follows.

The dissipative fluxes  $\mathbf{\delta\Pi}$ and $\mathbf{\delta J}$ in Eqs.~\eqref{eq:NS_moment} and \eqref{eq:NS_diffusion} are Gaussian stochastic processes with zero mean and variances that satisfy the fluctuation-dissipation relation \cite{Ortiz_de_Zarate_book} 
\begin{align}
        &\left\langle {\Pi}_{ij}(\mathbf{r}, t) {\Pi}_{kl}(\mathbf{r'}, t') \right \rangle  \\  &\qquad \quad= 2 k_BT \rho \nu \left( \delta_{ik}\delta_{jl}  + \delta_{il}\delta_{jk}\right)\delta^{(3)}(\mathbf{r} - \mathbf{r'})\delta(t - t'), \n
        \\ &\left\langle {J}_i(\mathbf{r}, t) {J}_j(\mathbf{r'}, t')\right \rangle   \label{FDT_Re}  
        = 2 k_BT \rho D  \chi \delta_{ij}\delta^{(3)}(\mathbf{r} - \mathbf{r'}
        )\delta(t - t'
        ), 
\end{align}
where $k_B$ is the Boltzmann constant and $T$ is the temperature. %
The considered  system is  infinite in the $y$ and $z$ directions, while $x \in [0, L]$ 
indicates the distance from one of the confining plates (walls). 
In an experimental realization of this setup,
the walls can be effectively constituted by surfaces of two larger spherical particles thoroughly immersed in the binary mixture (silica suspension). Such spherical particles, optically trapped at a certain distance from each other, can also be used for the simultaneous measurement of the fluctuation-induced pressure $p_{\rm Cas}$, as it displaces them from their equilibrium positions within the traps.

We note that,  in most liquids, the momentum diffusion is much faster than mass diffusion, i.e., $\nu/D \gg 1$ \cite{ortiz_de_zarate_non-equilibrium_2015}, which allows us to neglect 
the time derivative in Eq.~\eqref{eq:NS_moment}.
Taking the curl of the curl of the momentum balance in Eq.~\eqref{eq:NS_moment} and projecting it along $\mathbf{\hat{e}_z}$ renders 
\begin{align}
    \begin{bmatrix}
        (\mathbf{k}^2 - \partial_x^2)^2
,  & -k_{\mathrm{RO}}^4(k_y^2 -\partial^2_{ x} ) \\ 
        1, & i \omega/D  + (\mathbf{k}^2- \partial^2_{ x}) 
    \end{bmatrix}
    &
    \begin{bmatrix}
        \delta v_z(x, \mathbf{k}, \omega) \\ 
       (D/\nabla c) \delta c(x, \mathbf{k}, \omega)
    \end{bmatrix} \n \\ &
     = \mathbf{F}(x, \mathbf{k}, \omega), 
     \label{HD_Eq3}
\end{align}
where we performed the  Fourier transform in the $y$ and $z$  directions and in time, introducing the wavevector $\mathbf{k} = \left(0, k_y, k_z \right)$, and we denoted $\delta v_z = \mathbf{\hat{e}}_z \cdot \delta \mathbf{v}$. In the expression above, $\mathbf{F}(x, \mathbf{k}, \omega)$ is the stochastic forcing in the Fourier space, described in more detail in the SM.
In Eq.~\eqref{HD_Eq3} we introduced the roll-off 
wavevector defined as  (recall $\nabla c<0$) \cite{giavazzi_soft_matter_2016, Ortiz_de_Zarate_book}
\begin{equation}
    k_\mathrm{RO}^4 = -\frac{\beta g \nabla c}{D \nu},
    \label{Kro}
\end{equation}
where $\beta$ is given by Eq.~\eqref{eq:def-b}. $k_{\rm RO}$ describes the effect of gravity on the spectrum of fluctuations {of the concentration}: for $k>k_{\rm RO}$, the mean squared amplitude of the fluctuations scales
as $k^{-4}$, and the fluctuations relax by diffusion, whereas for $k< k_{\rm RO}$, the amplitude is reduced by buoyancy \cite{Ortiz_de_Zarate_book, giavazzi_soft_matter_2016}. We rederive this well-known fact  in the SM.

In the presence of the walls, the hydrodynamic fields $\delta v_z$ and $\delta c$ in Eq.~\eqref{HD_Eq3} are subject to BCs. In the considered case,   
we impose free BCs on the velocity fluctuations, i.e.,
\begin{align}
    \delta v_z(x, \mathbf{k}, \omega)|_{x = 0} = \delta v_z(x, \mathbf{k}, \omega)|_{x = L} = 0, \label{Free_BC_1} \\ 
    \partial^2_x\delta v_z(x, \mathbf{k}, \omega)|_{x = 0} = \partial^2_x\delta v_z(x, \mathbf{k}, \omega)|_{x = L} =0, 
     \label{Free_BC_2}
\end{align}
and Dirichlet BCs on the concentration fluctuations, i.e.,
\begin{equation}
    \delta c(x, \mathbf{k}, \omega)|_{x = 0} = \delta c(x, \mathbf{k}, \omega)|_{x = L} =0.
    \label{Dirichlet_BC}
\end{equation}
This specific choice allows us to express both $\delta c$ and $ \delta v_z$ as series of sine-modes along the $x$-direction,  with the corresponding  eigenvalues $(n\pi/L)$, as we discuss in more detail in the SM. 
Considering these \textit{model} BCs enables a semi-analytical derivation of the finite-size contribution to the NE pressure. 
Imposing the experimentally relevant BCs, i.e.,  Neumann BCs for $\delta c$ and no-slip BCs for $\delta v_z$ \cite{Ortiz_de_Zarate_2002} requires a more involved treatment {which goes beyond the scope of this analysis.} We describe the resulting issues in the SM.   

\section{The Casimir pressure}
When  $\nabla c \neq \nabla c_{\mathrm{gr}}$, 
the presence of a NE mass flux in the system gives rise to a coupling between the hydrodynamic modes, enhancing the fluctuations within the liquid. 
The resulting NE fluctuation-induced contribution to the pressure $p_{\rm NE}$ is given by \cite{ortiz_de_zarate_non-equilibrium_2015, kirkpatrick_ortiz_de_Zarate_2016, Kirkpatrick_Ortiz_de_zarate_2015_Cas}
\begin{align}    
    &p_{\mathrm{NE}}(L)  = \frac{1}{2} \mathcal{A} \int\frac{\mathrm{d}^2\mathbf{k}}{(2 \pi)^2}       S^{\rm e}(\mathbf{k}; L), \label{p_NE}  
\end{align}
where $\mathcal{A}$ is expressed in terms of 
thermodynamic quantities (see the SM), which, within the framework of fluctuating hydrodynamics, are approximated by their values averaged over the sample. 
The sign of the constant $\mathcal{A}$ is material-dependent; consequently, the sign of the fluctuation-induced pressure is also material-dependent, as discussed, e.g., in Ref.~\cite{Kirkpatrick_Ortiz_de_zarate_2016_Cas}. 
In Eq.~\eqref{p_NE}, $S^{\rm e}(\mathbf{k}; L)$ is the excess structure factor relative to the equilibrium, given by
\begin{align}
    & (2 \pi)^2 \delta^{(2)}(\mathbf{k} - \mathbf{k'})S^{\rm e}(\mathbf{k}; L) =  \n \\ &  \int_0^L \frac{\mathrm{d}x}{L}\int\frac{\mathrm{d} \omega \mathrm{d}\omega'}{(2 \pi)^2} \langle \delta c^*(x, \mathbf{k}, \omega) \delta c(x, \mathbf{k}', \omega')\rangle^{\rm e}_{\mathrm{NE}},
    \label{SF_def}
\end{align}
where $\langle \cdots \rangle^{\rm e}_{\mathrm{NE}} = \langle \cdots \rangle_{\mathrm{NE}} - \langle \cdots \rangle_{\mathrm{Eq}}$ is the difference between the non-equilibrium and equilibrium averages 
$\langle \cdot \rangle_{\mathrm{NE}}$ and $\langle \cdot \rangle_{\mathrm{Eq}}$, respectively.
From Eqs.~\eqref{HD_Eq3}, \eqref{p_NE}, and \eqref{SF_def} one eventually finds
\begin{align}
    p_{\mathrm{NE}}(L) = \frac{1}{2}\mathcal{A} \, S_E\left(\frac{\nabla c}{\nabla c_{\mathrm{gr}}} -1 \right)k^3_{\mathrm{RO}} \Psi(k_{\rm RO} L),
    \label{eq:p_NE_result}
\end{align}
where $S_E = \chi{k_BT}/{\rho}$ is the equilibrium static structure factor $S(\mathbf{k})$ in the hydrodynamic limit $\mathbf{k} \to 0$ \cite{giavazzi_soft_matter_2016}. 
We identify the dimensionless scaling variable  $X = k_{\rm RO}L$ as the one which controls the behavior of  the scaling function  $\Psi$.  For the 
model BCs given by  Eqs.~\eqref{Free_BC_1}, \eqref{Free_BC_2}, and \eqref{Dirichlet_BC}, the scaling function takes the form
\begin{equation}
    \Psi(X) = \frac{1}{X}\sum_{n=1}^{\infty} \int\frac{\mathrm{d}^{2}\mathbf{q}}{(2 \pi)^2} \frac{1}{1 + {q_n^6(X)}/{ q_{n\parallel}^2(X)}},
    \label{psi_deff}
\end{equation}
with $\mathrm{d}^{2}\mathbf{q} = \mathrm{d} q_y \mathrm{d} q_z$, where 
\begin{equation}   
     q_n^2(X) = \left(\frac{n \pi}{X}\right)^2 + q_y^2 + q_z^2 \quad \mbox{and}\quad q_{n\parallel}^2(X) = \left(\frac{n \pi}{X}\right)^2 + q_y^2.
     \label{momentum_n} 
\end{equation}
On the basis of Eq.~\eqref{eq:p_NE_result}, one can calculate the Casimir pressure $p_{\mathrm{Cas}}(L) $, defined as the finite-size contribution to the excess bulk pressure $p_{\mathrm{NE}}$ with respect to equilibrium, i.e., as 
\begin{equation}
p_{\mathrm{Cas}}(L) =  p_{\mathrm{NE}}(L) -  p_{\mathrm{NE}}(L \to \infty),
    \label{p_Cas}
\end{equation}
consistently with the literature concerning critical Casimir forces \cite{gambassi_critical_2024}. 
 \begin{figure*}
 \centering
\includegraphics[width=.65
\linewidth]{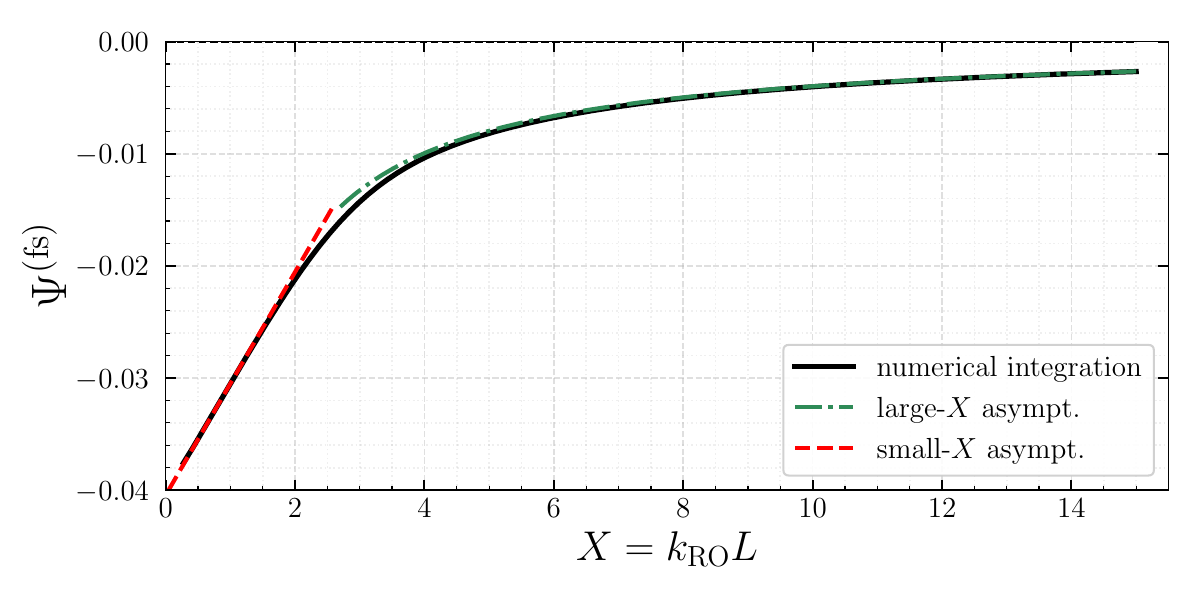}
\caption{ 
Scaling function $\Psi^{(\rm fs)}(X)$ of $p_{\rm Cas}$   given by Eqs.~\eqref{psi_deff} and \eqref{psi_fs_deff}. The black solid line corresponds to $\Psi^{(\rm fs)}(X)$  calculated numerically (starting from $X\simeq 0.3$),  while  its asymptotic behaviors given by Eq.~\eqref{ass_large} for $X\gg 1$ and by Eq.~\eqref{ass_small} for $X\ll 1$ are plotted with dashed green and red lines, respectively. The asymptotic expressions provide accurate approximation of the actual curve, except for $2\lesssim X \lesssim 4$.
} 
\label{python_plot}
\end{figure*} 
By using Eq.~\eqref{eq:p_NE_result} for $p_{\mathrm{NE}}(L)$, it is then straightforward to express $p_{\rm Cas}(L)$ as  in \cref{eq:result_anticipate},
where $\Psi^{\rm (fs)}$ is  the finite-size contribution to the scaling function defined as 
\begin{align}
    \Psi^{(\rm fs)}(X) =  \Psi(X) - \lim_{X \to \infty} \Psi(X).
    \label{psi_fs_deff}
\end{align}
In \cref{eq:result_anticipate}, we identify $k_{\rm RO}^{-1}$ as the  relevant length scale describing the rate of decay of the pressure with increasing separation $L$ between the plates. 
We  note that its value, given by Eq.~\eqref{Kro}, is independent of both the geometry and the BCs considered. 
Accordingly,  the scaling function $\Psi^{(\rm fs)}(X)$ in \cref{eq:result_anticipate} is the only part of the expression which depends on them. 

Figure~\ref{python_plot} shows $\Psi^{(\rm fs)}(X)$ as a function of $X$, obtained by numerically evaluating Eq.~\eqref{psi_deff} (see the SM).
The asymptotic behavior of $\Psi(X)$ for $X\gg 1$ can be determined analytically with the help of the Mellin transform \cite{Mellin}, finding 
\begin{align}
    \Psi(X) = \frac{1}{8\sqrt{2\pi}}\frac{\Gamma\left({7}/{4} \right)}{\Gamma\left({9}/{4} \right)} - \frac{1}{8 \pi X} +  \mathcal{O}(X^{-7/3}), 
         \label{ass_large}
\end{align}
where $\Gamma(x)$ denotes the Euler gamma function. We identify the first, $X$-independent term as the bulk contribution to the scaling function, which cancels out in $\Psi^{\rm (fs)}(X)$. 
We note that the second term is negative and hence, in the case of a positive thermodynamic prefactor $\mathcal{A}>0$, the resulting NE Casimir pressure is attractive, while it is repulsive for $\mathcal{A}<0$. 
Moreover, for $X \gg 1$, i.e., $L \gg k^{-1}_{\rm RO}$, the finite-size contribution to the scaling function and the resulting  Casimir pressure vanish algebraically as $\sim L^{-1}$. Note that this decay is the same as the one predicted in Ref.~\cite{Aminov_2015} for the NE Casimir force in a simple driven diffusive system in the absence of gravity, with the density gradient parallel to the confining plates. There, the sign of the NE Casimir pressure depends on the dynamics considered. 
In the opposite case $X \ll 1$, we find
\begin{equation}
    \Psi(X) =  \frac{X}{32 \pi} + \mathcal{O}(X^5).
 \label{ass_small}
\end{equation}
The asymptotic behaviors in Eqs.~\eqref{ass_large} and \eqref{ass_small} provide accurate approximations of the numerical result presented in \cref{python_plot}, except for the region with $2 \lesssim X \lesssim 4$.

\subsection{Zero-gravity limit}
In the absence of gravity, the roll-off wavevector vanishes, i.e., $k_{\rm RO} \to 0 $ (see \cref{Kro}) and thus $X \to 0$. Correspondingly, the NE contribution to the pressure in Eq.~\eqref{eq:p_NE_result} takes the form

\begin{align}
    p_{\mathrm{NE}}(L) &= \mathcal{A}\frac{k_BT}{2 \rho \nu D } \left[1 - \frac{\nabla c_{\mathrm{gr}}}{\nabla c} \right](\nabla c)^2L\frac{\Psi(X)}{X}  \n \\ & 
    \quad\xrightarrow[g \to 0]{}  \mathcal{A}\frac{k_BT}{2 \rho \nu D }(\nabla c)^2 L\frac{1}{32 \pi },
\end{align}
 where we used that, according to Eq.~\eqref{ass_small}, $\lim_{X \to 0}{\Psi(X)}/{X} = {1}/{(32 \pi)}$.  As we discuss in the SM,  in the case of the gradient perpendicular  to the walls,  
 the function $\Psi_\bot$ corresponding to $\Psi$ satisfies
 $\lim_{X \to 0}{\Psi_{\perp}(X)}/{X} = {1}/{(48 \pi)}$, recovering the analogous mode-coupling result for one-component fluids subject to a temperature gradient with the same  boundary conditions (free/Dirichlet) as those considered here, 
 reported in Refs.~\cite{Kirkpatrick_Giant_2013, Kirkpatrick_Ortiz_de_zarate_2014, kirkpatrick_ortiz_de_Zarate_2016}.  
 Note that in the  zero-gravity limit,
 $p_{\mathrm{NE}}(L) = \mathrm{const.} \times (\nabla c)^2 L$, irrespective of the  geometry considered, as predicted in 
 Refs.~\cite{Kirkpatrick_Giant_2013, Kirkpatrick_Ortiz_de_zarate_2014, 
Kirkpatrick_Ortiz_de_zarate_2015_Cas, kirkpatrick_ortiz_de_Zarate_2016, ortiz_de_zarate_non-equilibrium_2015, Kirkpatrick_Ortiz_de_zarate_2016_Cas}. 

\subsection{In colloidal suspensions}
\label{subsec_colloid}
If the binary liquid mixture consists of a colloidal suspension, with tiny colloidal particles as the solute in the mixture and, e.g., water as the solvent, we can assume that the osmotic susceptibility $\chi$ is given by \cite{Russel_book}  
\begin{align}
    \chi^{-1} = \frac{k_BT}{m_p c}Z(\phi),
    \label{eq:osmotic_colloid}
\end{align}
where $m_p$ is the mass of a single silica particle, $\phi = ({c \rho})/\rho_p$ is the volume fraction of the solute, and $\rho_p$ is the material density of colloidal particles. The dimensionless compressibility factor $Z(\phi)$ above quantifies the deviation from the case of an ideal mixture, for which a classical result due to van't Hoff \cite{Hoff} renders $Z(\phi) \equiv 1$. Moreover, as we show in the SM, assuming that the colloidal particles are incompressible,
the thermodynamic prefactor $\mathcal{A}\, S_E $ in \cref{eq:result_anticipate} simplifies, and the expression for the Casimir pressure takes the particularly simple form
\begin{align}
    p_{\mathrm{Cas}}(L) = \frac{1}{2} k_BTs\left(\frac{\nabla c}{\nabla c_{\mathrm{gr}}} -1 \right)k^3_{\mathrm{RO}} \Psi^{\rm (fs)}(k_{\rm RO} L),
    \label{eq:p_NE_result_colloid}
\end{align}
where 
\begin{align}
    s = \frac{\mathrm{d} \log Z(\phi)}{\mathrm{d} \log \phi}.
    \label{eq:s_def}
\end{align}
Note that $s=0$ for an ideal mixture:
accordingly, no NE fluctuation-induced pressure is observed, and the presence of the Casimir pressure can be regarded as a consequence of non-ideal mixing in the mixture.

For a dense solution of LUDOX TMA silica particles dispersed in water which diffuses into an overlaid layer of pure water, the roll-off wavevector was measured \cite{giavazzi_soft_matter_2016} and turned out to be $k_{\rm RO} \simeq 0.1 \, \mu m^{-1}$. 
Moreover, the experimental data of 
Ref.~\cite{giavazzi_soft_matter_2016} allow us to estimate the value of the parameter $s$ defined in  \cref{eq:s_def}, which turns out to be $s = 1.65 \pm 0.10$, see the SM for details. 

From Eqs.~\eqref{psi_deff} and \eqref{psi_fs_deff}, instead, one finds $\Psi^{\rm(fs)}(1) \simeq -0.03$. Accordingly, in such a system, the Casimir pressure $p_{\rm Cas}$ is negative, i.e., the resulting force  is attractive. The estimates of the order of magnitude of $|p_{\rm Cas}|$ and of the  force acting on a $1 \mbox{mm}^2$ plate are reported in  \cref{tab:values}. 
In an experiment with two spherical particles trapped by optical tweezers at a surface-to-surface distance $L$ of the order of $L \approx k_{\rm RO}^{-1}$, the force acting on them can be estimated as $F \simeq p_{\rm Cas}R_c^2$, where $R_c$ is the radius of the larger colloids.
Assuming for $R_c$ and for the concentration gradient values which are within experimental reach, i.e., $R_c \simeq 10\, \mu \mbox{m}$ and $(\nabla c/\nabla c_{\rm gr}-1) \simeq 10^2$,  the force turns out to be of the order of a few femtonewtons.
Measuring fN forces in liquid environments in optical-tweezer experiments is challenging but not impossible. 
Using soft traps, long averaging times, differential, oscillatory or lock-in detection allowed, in fact,  to reach the necessary resolution in experiments, e.g., on single DNA molecules in aqueous solutions \cite{Meiners} and electrostatic (double-layer) forces between polymer microspheres \cite{Sainis}. 
In this regime, van der Waals surface forces  have been measured by modulating an optically trapped polystyrene probe near a silica surface in a liquid \cite{Kundu}. More recently, sub-femtonewton force resolution has been achieved for a dielectric microsphere optically trapped in water near a dielectric interface \cite{Liu}, and at the thermal-noise limit using high-refractive-index silicon nanospheres in ambient-temperature water, where forces as low as $\simeq 0.3\, \mbox{fN}$ were measured \cite{Kostarev}. Accordingly, forces of a few fN lie within the capabilities of optimized optical-tweezer force spectroscopy.

\begin{table}
    \centering
\begin{center}
\begin{tabular}{ ||c | c | c|| 
}  \hline\hline
   $\nabla c/\nabla c_{\rm gr}-1$& $|p_{\rm Cas}|$ [Pa] & $f$ [pN] 
   \\
  \hline \hline
  $10$ & $\simeq  1 \times10^{-6}$ & $ \simeq 1$
  \\ 
  $10^2$ &$ \simeq 1\times 10^{-5}$ & $ \simeq 10$ 
  \\ 
  
   $10^3$& $\simeq 1 \times  10^{-4}$ & $ \simeq 10^2$ 
   \\ \hline\hline
\end{tabular}
\end{center}
    \caption{Estimated values of the NE Casimir pressure $p_{\rm Cas}$ and of the corresponding force $f$ acting on the confining surfaces of a film of thickness $L = 10\,\mu\mbox{m}$ and surface area $1\, \mbox{mm}^2$ 
    due to a dense aqueous colloidal suspension (LUDOX TMA silica particles dispersed in water) diffusing into an overlying layer of pure water, a system investigated in Ref.~\cite{giavazzi_soft_matter_2016}.}
    \label{tab:values}
\end{table} 

\section{Conclusions}
We predicted the non-equilibrium Casimir pressure  $p_{\mathrm{Cas}}(L)$ acting on the walls of a three-dimensional slab  of thickness $L$ and arising from the confinement of hydrodynamic fluctuations in an isothermal binary  liquid mixture in the presence of a constant non-equilibrium concentration gradient $\bm{\nabla} c$ parallel to the gravitational acceleration and the slab walls. We considered the case of the gradient parallel to the walls and co-directional with the gravitational acceleration in a  system far from convective instability. Importantly, we assumed that the liquid 
fills not only the space between the walls, but also outside so that $p_{\mathrm{Cas}}(L)$ gives the net pressure acting on each wall. 

The sign and the magnitude of $p_{\mathrm{Cas}}(L)$ depend on a thermodynamic constant $\mathcal{A}$ and the pressure vanishes at equilibrium. Moreover, $p_{\mathrm{Cas}}(L)$  can be expressed in terms of a dimensionless scaling function $\Psi^{\rm (fs)}(X)$ of a scaling variable $X = k_{\rm RO}L$, where $k_{\rm RO}$  is the roll-off wavevector (see Eq.~\eqref{Kro}) describing the suppression of thermal fluctuations due to the presence of gravity. 
For large $L$, the pressure vanishes algebraically $\sim L^{-1}$.  
The pressure was derived for the case of model boundary conditions, i.e., free boundary conditions for the velocity field fluctuations, and Dirichlet boundary conditions for the concentration fluctuations. A similar derivation for a different set of boundary conditions, e.g., no-slip for the velocity field and Neumann for the concentration,  remains an open problem. Finally, we found a simplified expression describing the Casimir pressure in colloidal suspensions. In  a light-scattering experiment in a dense aqueous colloidal suspension, i.e., LUDOX TMA silica particles dispersed in water, diffusing into an overlying layer of pure water \cite{giavazzi_soft_matter_2016},  the roll-off wavevector was measured to be $k_{\rm RO} \approx 0.1\, \mu \mbox{m}^{-1}$. 
We estimate the magnitude of the force acting on spherical particles immersed in such systems and trapped optically at a distance $\simeq k_{\rm RO}^{-1}$ from each other to be of the order of a few  femtonewtons.

\providecommand{\newblock}{}

\onecolumngrid
\clearpage


\section*{Supplementary material}

# Supplemental Material
# for "Gravity-controlled non-equilibrium Casimir pressure in a binary liquid mixture"

Marcin Piotr Pruszczyk,[1,2] Roberto Cerbino,[3] and Andrea Gambassi[1,2]
[1] *SISSA — International School for Advanced Studies, via Bonomea 265, 34136 Trieste, Italy*
[2] *INFN, via Bonomea 265, 34136 Trieste, Italy*
[3] *Faculty of Physics, University of Vienna, Boltzmanngasse 5, 1090 Vienna, Austria*


## CONTENTS



In this Supplemental Material, we provide the analytical derivations and describe the numerical computations underlying the results presented in the main text. In particular, in Sec. S.I, we discuss the thermodynamic constant $\mathcal{A}$ and its simplification for aqueous colloidal suspensions. In Sec. S.II, we provide the details regarding the hydrodynamic equations in the Boussinesq approximation. In Sec. S.III, we derive the structure factor in the case of the model boundary conditions (BCs) discussed in the present work. In Sec. S.IV, we elaborate on the difficulties arising from considering more realistic BCs and in particular from imposing (i) no-slip BCs on the velocity fluctuations or (ii) Neumann BCs on the concentration fluctuations while simultaneously imposing free BCs on the velocity fluctuations. In Sec. S.V, we describe the numerical calculation of the scaling function of the non-equilibrium pressure based on the Poisson resummation formula, and discuss its limitations. In Sec. S.VI, by means of the Mellin transform, we derive analytically the asymptotic behavior of the scaling function, both in the case of the gradient parallel and perpendicular to the confining walls. Finally, in Sec. S.VII we provide an estimate of the logarithmic derivative $s$ of the compressibility factor in Eq. (26) of the main text, which determines the amplitude of the Casimir pressure $p_{\mathrm{Cas}}$ in Eq. (25). This estimate is based on the experimental data of Ref. [13].

## S.I. THE THERMODYNAMIC PREFACTOR $\mathcal{A}$

The thermodynamic prefactor $\mathcal{A}$ introduced in Eq. (14) in the main text is given by [1]

$$\mathcal{A} = -\frac{\rho(\gamma-1)}{\alpha T}\left[\chi^{-1} - T\left(\frac{\partial \chi^{-1}}{\partial T}\right)_{p,c} - \frac{\rho \mathrm{c}_p}{\alpha}\left(\frac{\partial \chi^{-1}}{\partial p}\right)_{T,c}\right], \tag{S1}$$

where $\rho$ is the total mass density of the mixture, $\gamma = \mathrm{c}_p/\mathrm{c}_V$ is the ratio of the isobaric and isochoric heat capacities, $\mathrm{c}_p$ and $\mathrm{c}_V$, respectively, $\alpha = -\rho^{-1}(\partial\rho/\partial T)_{p,c}$ is the thermal expansion coefficient, $T$ is the temperature of the mixture,

and $\chi = (\partial c/\partial \mu)_{p,T}$ is the osmotic susceptibility. (Despite the similar notation, these heat capacities $c_p$ and $c_V$ are unrelated to the solute concentration $c$ and should not be confused with it.) In the framework of fluctuating hydrodynamics, all these thermodynamic quantities are assumed to be given by their average values across the sample. We note that the sign of $\mathcal{A}$ — and thus of the corresponding fluctuation-induced force $\propto \mathcal{A}$ — depends on the involved substances, as shown, e.g., in Ref. [1]. We now use the thermodynamic relation [2]

$$c_p - c_V = \frac{\alpha^2 T}{\rho \kappa_T}, \tag{S2}$$

where $\kappa_T = \rho^{-1}(\partial \rho/\partial p)_{c,T}$ is the isothermal compressibility, to rewrite the overall factor $-\rho(\gamma - 1)/(\alpha T)$ in Eq. (S1) as $-\alpha/(\kappa_T c_V)$. As discussed in the main text, for a colloidal suspension, we can assume that [3]

$$\chi^{-1} = \frac{k_B T}{m_p c} Z(\phi), \tag{S3}$$

where $m_p$ is the mass of a single silica particle, $\phi = (c\rho)/\rho_p$ is the volume fraction of the solute, and $\rho_p$ is the material density of colloidal particles. In fact, this relation encompasses the van't Hoff relation for an ideal mixture [4], which corresponds to $Z(\phi) \equiv 1$. Accordingly, the corrections to the ideal behavior are described by the dimensionless constant $Z$. Additionally, assuming that the solute particles are incompressible, i.e., that the dependence of $\phi$ on $p$ and $T$ at fixed concentration $c$ enters solely through $\rho$, and using the definitions of the thermal expansion coefficient $\alpha = -\rho^{-1}(\partial \rho/\partial T)_{p,c}$ and the isothermal compressibility $\kappa_T = \rho^{-1}(\partial \rho/\partial p)_{c,T}$, we obtain

$$\left(\frac{\partial \phi}{\partial T}\right)_{p,c} = -\alpha \phi, \tag{S4}$$

$$\left(\frac{\partial \phi}{\partial p}\right)_{T,c} = \kappa_T \phi. \tag{S5}$$

Under these assumptions, a straightforward calculation utilizing the relation in Eq. (S2) and the expressions for the partial derivatives above renders

$$\mathcal{A} = \rho \frac{k_B T}{m_p c} \phi \frac{\mathrm{d} Z(\phi)}{\mathrm{d}\phi} = \rho \chi^{-1} \frac{\phi}{Z(\phi)} \frac{\mathrm{d} Z(\phi)}{\mathrm{d}\phi} = \rho \chi^{-1} \frac{\mathrm{d} \log Z(\phi)}{\mathrm{d} \log \phi}. \tag{S6}$$

Recalling the definition of the static structure factor $S_E$ given below Eq. (16) in the main text, i.e.,

$$S_E = k_B T \chi/\rho, \tag{S7}$$

we can use the previous relation to write

$$\mathcal{A} S_E = k_B T s, \tag{S8}$$

where $s$ is the logarithmic derivative of the compressibility factor introduced in Eq. (26) of the main text. This expression allows us to simplify the one for the NE Casimir force in Eq. (1), which eventually takes the form reported in Eqs. (25) and (26) therein. In, c.f., Sec. S.VII we provide an estimate of $s$ based on the available experimental data.

## S.II. SIMPLIFICATION OF THE FLUCTUATING-HYDRODYNAMICS EQUATION

In this section, we discuss the simplification of the fluctuating hydrodynamic equations in the Boussinesq approximation [5]. Namely, we justify the omission of the time derivative in the momentum balance in Eq. (3) and perform the Fourier transform in the $y$ and $z$ directions of the fluctuating quantities $\delta v_z$ and $\delta c$, allowing for the derivation of the structure factor in the case of mode BCs in Section S.III.

Taking curl of the curl of the momentum balance in Eq. (3) in the main text and projecting it along $\hat{\mathbf{e}}_\mathbf{z}$ renders

$$\begin{bmatrix} (\nu^{-1}\partial_t \nabla^2 - \nabla^4), & -k_{\mathrm{RO}}^4(\partial_x^2 + \partial_y^2) \\ 1, & (D^{-1}\partial_t - \nabla^2) \end{bmatrix} \begin{bmatrix} \delta v_z(\mathbf{r},t) \\ \frac{D}{\nabla c}\delta c(\mathbf{r},t) \end{bmatrix} = \mathbf{F}(\mathbf{r},t), \tag{S9}$$

where we used the definition of the roll-off wavevector in Eq. (10) in the main text. The stochastic forcing on the r.h.s. of the Eq. (S9) is of the form

$$\mathbf{F}(\mathbf{r},t) = \frac{1}{\rho} \begin{bmatrix} \nu^{-1} [\boldsymbol{\nabla} \times (\boldsymbol{\nabla} \times (\boldsymbol{\nabla} \cdot \delta \boldsymbol{\Pi}(\mathbf{r},t)))]_z \\ (\nabla c)^{-1} [-\boldsymbol{\nabla} \cdot \delta \mathbf{J}(\mathbf{r},t)] \end{bmatrix}, \tag{S10}$$

where $\delta\mathbf{\Pi}$ and $\delta\boldsymbol{J}$ are the noise terms introduced in Eqs. (3) and (4) in the main text, satisfying the fluctuation dissipation relation expressed in Eqs. (7) and (8) therein. As discussed above Eq. (9) in the main text, the time derivative acting on the velocity fluctuation in Eq. (S9) can actually be neglected. In order to see why this is the case, we introduce the dimensionless quantities

$$\tilde{x} = x/L, \quad \tilde{t} = Dt/L^2, \quad \delta\tilde{c} = \delta c/(\nabla c\, L), \quad \text{and} \quad \delta\tilde{v}_z = \delta v_z L/D. \tag{S11}$$

In terms of these dimensionless variables, the l.h.s. of Eq. (S9) takes the form

$$\begin{bmatrix} -\mathrm{Sc}^{-1}\partial_{\tilde{t}}\widetilde{\nabla}^2 + \widetilde{\nabla}^4, & X^4\left(\partial_{\tilde{x}}^2 + \partial_{\tilde{y}}^2\right) \\ 1, & \left(\partial_{\tilde{t}} - \widetilde{\nabla}^2\right) \end{bmatrix} \begin{bmatrix} \delta\tilde{v}_z(\tilde{\mathbf{r}}, \tilde{t}) \\ \delta\tilde{c}(\tilde{\mathbf{r}}, \tilde{t}) \end{bmatrix} \tag{S12}$$

where $\widetilde{\nabla} = L\nabla$, $\mathrm{Sc} = \nu/D$ is the so-called Schmidt number [5], and $X = k_{\mathrm{RO}}L$ is the scaling variable entering the scaling function. In most liquids it turns out that $\mathrm{Sc} \gg 1$ [6], which allows us to neglect the terms $\propto \mathrm{Sc}^{-1}$ and thus the time derivative acting on the velocity field fluctuation in Eq. (S9).

Additionally, we notice that the considered system is invariant under spatial translation along the $y$ and $z$ directions, which are parallel to the confining surfaces. Accordingly, it is convenient to express the fluctuations $\delta v_z$ and $\delta c$ in terms of their Fourier transforms

$$\delta v_z(\mathbf{r}, t) = \int \frac{\mathrm{d}\omega}{2\pi} \int \frac{\mathrm{d}^2\mathbf{k}}{(2\pi)^2} e^{i\omega t - i\mathbf{k}\cdot\mathbf{r}}\, \delta v_z(x, \mathbf{k}, t), \tag{S13}$$

$$\delta c(\mathbf{r}, t) = \int \frac{\mathrm{d}\omega}{2\pi} \int \frac{\mathrm{d}^2\mathbf{k}}{(2\pi)^2} e^{i\omega t - i\mathbf{k}\cdot\mathbf{r}}\, \delta c(x, \mathbf{k}, t), \tag{S14}$$

with $\mathbf{k} = (0, k_y, k_z)$. The inverses of these transforms are

$$\delta v_z(x, \mathbf{k}, t) = \int \mathrm{d}t \int \mathrm{d}y \int \mathrm{d}z\, e^{-i\omega t + i\mathbf{k}\cdot\mathbf{r}}\, \delta v_z(\mathbf{r}, t), \tag{S15}$$

$$\delta c(x, \mathbf{k}, t) = \int \mathrm{d}t \int \mathrm{d}y \int \mathrm{d}z\, e^{-i\omega t + i\mathbf{k}\cdot\mathbf{r}}\, \delta c(\mathbf{r}, t), \tag{S16}$$

where $\mathbf{r} = (x, y, z)$. This allows us to cast Eq. (S9) in the form of Eq. (9) in the main text, where $\mathbf{F}(x, \mathbf{k}, \omega)$ is the Fourier transform of the stochastic forcing in Eq. (S10).

## S.III. DERIVATION OF THE STRUCTURE FACTOR FOR MODEL BOUNDARY CONDITIONS

In the presence of spatial confinement, boundary conditions (BCs) are imposed on fluctuations by the walls of the slab. In the model case described in the main text, we consider free BCs on the velocity fluctuations reported in Eqs. (11) and (12) therein. Accordingly, $\delta v_z$ can be expressed as a series

$$\delta v_z(x, \mathbf{k}, \omega) = \sum_{n=1}^{\infty} \delta v_z^{(n)}(\mathbf{k}, \omega)\sqrt{2}\sin\left(\frac{n\pi x}{L}\right). \tag{S17}$$

In addition, we impose Dirichlet BCs on the concentration fluctuations, defined in Eq. (13) in the main text, which allows one to express $\delta c$ as

$$\delta c(x, \mathbf{k}, \omega) = \sum_{n=1}^{\infty} \delta c^{(n)}(\mathbf{k}, \omega)\sqrt{2}\sin\left(\frac{n\pi x}{L}\right). \tag{S18}$$

We note that

$$\frac{2}{L}\int_0^L \mathrm{d}x\, \sin\left(\frac{n\pi x}{L}\right)\sin\left(\frac{m\pi x}{L}\right) = \delta_{nm}, \tag{S19}$$

where $\delta_{nm}$ is the Kronecker delta. Accordingly, the modes $\propto \sin(n\pi x/L)$ constitute an orthonormal basis with respect to the scalar product

$$(f|g) = \frac{1}{L}\int_0^L \mathrm{d}x f(x)g(x). \tag{S20}$$

Consequently, utilizing this product and the Fourier transform introduced in Eqs. (S15) and (S16), the fluctuating hydrodynamics equations (S9) can be projected onto a system of algebraic equations of the form

$$\begin{bmatrix} k_n^4, & -k_{\mathrm{RO}}^4 k_{n\parallel}^2 \\ 1, & i\frac{\omega}{D} + k_n^2 \end{bmatrix} \begin{bmatrix} \delta v_z^{(n)}(\mathbf{k},\omega) \\ \frac{D}{\nabla c}\delta c^{(n)}(\mathbf{k},\omega) \end{bmatrix} = \begin{bmatrix} F_1^{(n)}(\mathbf{k},\omega) \\ F_2^{(n)}(\mathbf{k},\omega) \end{bmatrix} \qquad \text{for} \quad n \geq 1, \tag{S21}$$

where the wavevectors are given by

$$k_n^2 = \left(\frac{n\pi}{L}\right)^2 + k_y^2 + k_z^2, \tag{S22}$$

$$k_{n\parallel}^2 = \left(\frac{n\pi}{L}\right)^2 + k_y^2, \tag{S23}$$

and the projection $F_i^{(n)}(\mathbf{k},\omega)$ of the stochastic forcing by

$$F_i^{(n)}(\mathbf{k},\omega) = \frac{1}{L}\int_0^L \mathrm{d}x\sqrt{2}\sin\left(\frac{n\pi x}{L}\right) F_i(x,\mathbf{k},\omega). \tag{S24}$$

We apply the orthonormality condition in Eq. (S19) to the series representation of $\delta c$ in Eq. (S18), which renders

$$\int_0^L \frac{\mathrm{d}x}{L}\int \frac{\mathrm{d}\omega\mathrm{d}\omega'}{(2\pi)^2}\langle \delta c^*(x,\mathbf{k},\omega)\delta c(x,\mathbf{k}',\omega')\rangle_{\mathrm{NE}} = \sum_{n=1}^{\infty}\int \frac{\mathrm{d}\omega\mathrm{d}\omega'}{(2\pi)^2}\langle \delta c^{(n)*}(\mathbf{k},\omega)\delta c^{(n)}(\mathbf{k}',\omega')\rangle_{\mathrm{NE}}. \tag{S25}$$

The correlation functions of fluctuations such as those on the r.h.s. of the previous equation can be determined by inverting the relation in Eq. (S21), which yields

$$\begin{bmatrix} \delta v_z^{(n)}(\mathbf{k},\omega) \\ \frac{D}{\nabla c}\delta c^{(n)}(\mathbf{k},\omega) \end{bmatrix} = \frac{1}{k_n^4(i\omega/D + k_n^2) + k_{\mathrm{RO}}^4 k_{n\parallel}^2} \begin{bmatrix} i\frac{\omega}{D} + k_n^2, & k_{\mathrm{RO}}^4 k_{n\parallel}^2 \\ -1, & k_n^4 \end{bmatrix} \begin{bmatrix} F_1^{(n)}(\mathbf{k},\omega) \\ F_2^{(n)}(\mathbf{k},\omega) \end{bmatrix}, \tag{S26}$$

from which one can easily read off $\delta c^{(n)}(\mathbf{k},\omega)$ and the resulting correlation

$$\begin{aligned} \langle \delta c^{(n)*}(\mathbf{k},\omega)\delta c^{(n)}(\mathbf{k}',\omega')\rangle_{\mathrm{NE}} = \left(\frac{\nabla c}{D}\right)^2 &\frac{1}{k_n^4(-i\omega/D + k_n^2) + k_{\mathrm{RO}}^4 k_{n\parallel}^2}\,\frac{1}{k_n'^4(i\omega'/D + k_n'^2) + k_{\mathrm{RO}}^4 k_{n\parallel}'^2}\times \\ &\left[\left\langle F_1^{(n)*}(\mathbf{k},\omega)F_1^{(n)}(\mathbf{k}',\omega')\right\rangle + k_n^4 k_n'^4\left\langle F_2^{(n)*}(\mathbf{k},\omega)F_2^{(n)}(\mathbf{k}',\omega')\right\rangle\right]. \end{aligned} \tag{S27}$$

The fluctuation dissipation relation in Eqs. (7) and (8) in the main text renders the correlation of the projected stochastic forcing (see Eqs. (S10) and (S24)) in the form

$$\left\langle F_1^{(n)*}(\mathbf{k},\omega)F_1^{(n)}(\mathbf{k}',\omega')\right\rangle = (2\pi)\delta(\omega-\omega')(2\pi)^2\delta^{(2)}(\mathbf{k}-\mathbf{k}')\frac{2k_BT}{\rho\nu L}k_n^4 k_{n\parallel}^2, \tag{S28}$$

$$\left\langle F_2^{(n)*}(\mathbf{k},\omega)F_2^{(n)}(\mathbf{k}',\omega')\right\rangle = (2\pi)\delta(\omega-\omega')(2\pi)^2\delta^{(2)}(\mathbf{k}-\mathbf{k}')\frac{2k_BT\chi D}{\rho(\nabla c)^2 L}k_n^2, \tag{S29}$$

where we used the expression for the inverse Fourier transform in Eq. (S15), and where $k_n^2$ and $k_{n\parallel}^2$ are given by Eqs. (S22) and (S23), respectively. The equations above allow us to cast Eq. (S27) in the form

$$\begin{aligned} \langle \delta c^{(n)*}(\mathbf{k},\omega)\delta c^{(n)}(\mathbf{k}',\omega')\rangle_{\mathrm{NE}} =& \frac{2k_BT\chi}{\rho L}\delta(\omega-\omega')(2\pi)^2\delta^{(2)}(\mathbf{k}-\mathbf{k}') \\ &\times \frac{Dk_n^2}{[\omega + iDk_n^2(1+(k_{n\parallel}^2 k_{\mathrm{RO}}^4/k_n^6))][\omega - iDk_n^2(1+(k_{n\parallel}^2 k_{\mathrm{RO}}^4/k_n^6))]}\left[1 + \frac{\nabla c}{\nabla c_{\mathrm{gr}}}\frac{k_{n\parallel}^2 k_{\mathrm{RO}}^4}{k_n^6}\right]. \end{aligned} \tag{S30}$$

A straightforward integration over $\omega$ and $\omega'$ yields

$$\int \frac{\mathrm{d}\omega\mathrm{d}\omega'}{(2\pi)^2}\langle \delta c^{(n)*}(\mathbf{k},\omega)\delta c^{(n)}(\mathbf{k}',\omega')\rangle_{\mathrm{NE}} = (2\pi)^2\delta^{(2)}(\mathbf{k}-\mathbf{k}')\frac{k_BT\chi}{\rho}\frac{1}{L}\left[1 + \left(\frac{\nabla c}{\nabla c_{\mathrm{gr}}} - 1\right)\frac{1}{1 + k_n^6/(k_{\mathrm{RO}}^4 k_{n\parallel}^2)}\right]. \tag{S31}$$

We note that the first term in the square brackets corresponds to the value of the integral at equilibrium, and thus the second term represents the excess NE contribution, which vanishes for $\nabla c = \nabla c_{\text{gr}}$. We recall the expression for the static structure factor $S_E$ in Eq. (S7), which is the overall amplitude in Eq. (S31), and use Eq. (15) in the main text to write

$$\int_0^L \frac{\mathrm{d}x}{L} \int \frac{\mathrm{d}\omega \mathrm{d}\omega'}{(2\pi)^2} \langle \delta c^*(x,\mathbf{k},\omega)\delta c(x,\mathbf{k}',\omega')\rangle^{\text{e}}_{\text{NE}} = (2\pi)^2 \delta^{(2)}(\mathbf{k}-\mathbf{k}') S^{\text{e}}(\mathbf{k};L), \tag{S32}$$

where

$$S^{\text{e}}(\mathbf{k};L) = S_E \left(\frac{\nabla c}{\nabla c_{\text{gr}}} - 1\right) \frac{1}{L} \sum_{n=1}^{\infty} \frac{1}{1 + k_n^6/(k_{\text{RO}}^4 k_{n\|}^2)}. \tag{S33}$$

Note that the summand in the expression above, as a function of $k_n$, vanishes as $\sim k_n^{-4}$ upon increasing $k_n$ beyond $k_n > k_{\text{RO}}$, while it approaches a finite value for $k_n < k_{\text{RO}}$. Integrating over the momenta in the expression above straightforwardly leads to

$$\int \frac{\mathrm{d}^2\mathbf{k}}{(2\pi)^2} S^{\text{e}}(\mathbf{k};L) = S_E \left(\frac{\nabla c}{\nabla c_{\text{gr}}} - 1\right) k_{\text{RO}}^3 \frac{1}{k_{\text{RO}}^3 L} \sum_{n=1}^{\infty} \int \frac{\mathrm{d}^2\mathbf{k}}{(2\pi)^2} \frac{1}{1 + k_n^6/(k_{\text{RO}}^4 k_{n\|}^2)}. \tag{S34}$$

Now, changing the variables according to $\mathbf{k} \mapsto \mathbf{q} = \mathbf{k}/k_{\text{RO}}$ renders the expression for the scaling function $\Psi(X)$ (recall $X = k_{\text{RO}} L$), which is given by Eqs. (17), and (18) of the main text.

## S.IV. DIFFICULTIES WITH OTHER BOUNDARY CONDITIONS

In the main text and in the section above, we considered model boundary conditions, which allow one to evaluate the finite-size contribution to the structure factor and the resulting scaling function, as discussed, c.f., in Secs. S.V and S.VI. Note, however, that the boundary conditions considered here differ from those realized experimentally. They are nevertheless adopted because they permit an analytical determination of the finite-size contribution, which is not feasible for the experimental boundary conditions, as explained below.

### A. No-slip BCs for the velocity fluctuations

The physically relevant BCs imposed on the velocity field fluctuation are the so-called no-slip BCs [6]

$$\delta v_z(x,\mathbf{k},\omega)|_{x=0} = \delta v_z(x,\mathbf{k},\omega)|_{x=L} = 0, \tag{S35}$$

$$\partial_x \delta v_z(x,\mathbf{k},\omega)|_{x=0} = \partial_x \delta v_z(x,\mathbf{k},\omega)|_{x=L} = 0. \tag{S36}$$

The eigenfunctions $u_n(x)$ of the operator acting on the velocity field in Eq. (9) in the main text (the top-left entry of the matrix therein), i.e., the functions satisfying

$$(\mathbf{k}^2 - \partial_x^2)^2 u_n(x) = (\mathbf{k}^2 + \kappa_n)^2 u_n(x), \tag{S37}$$

and the BCs in Eqs. (S35) and (S36) can be constructed in closed form, but the corresponding eigenvalues $\kappa_n$ are only implicitly defined as the solutions of a transcendental equation, see Ref. [7] for the details. Explicit expressions for $\kappa_n$ are not known and this does not allow one to derive the finite-size contribution using methods such as the Poisson resummation, described in Sec. S.V, and Mellin transform, described in Sec. S.VI.

### B. Neumann BCs for the concentration fluctuations

For the concentration fluctuations, the physically relevant BCs are the Neumann ones [1]:

$$\partial_x \delta c(x,\mathbf{k},\omega)|_{x=0} = \partial_x \delta c(x,\mathbf{k},\omega)|_{x=L} = 0, \tag{S38}$$

for which $\delta c$ can be expressed as a series

$$\delta c(x,\mathbf{k},\omega) = \sum_{n=0}^{\infty} \delta c^{(n)}(\mathbf{k},\omega) u_n(x), \tag{S39}$$

with the eigenfunctions $u_n(x)$ given by

$$u_n(x) = \begin{cases} 1 & \text{for} \quad n = 0, \\ \sqrt{2}\cos\left(\frac{n\pi x}{L}\right) & \text{for} \quad n \neq 0. \end{cases} \tag{S40}$$

Below, we distinguish matrices/operators acting on the space of modes labeled by $n = 0, 1, \ldots$ from column vectors in that same space by denoting the former with a double underline (e.g., $\underline{\underline{K}}^2$) and the latter with a single underline (e.g., $\underline{v}_z$).

In the case of free BCs imposed on the velocity-field fluctuations (see Eqs. (11) and (12) in the main text) and Neumann BCs imposed on the concentration fluctuations, the differential equation (S9) becomes the algebraic equation

$$\begin{bmatrix} \underline{\underline{K}}^4 & -k_{\mathrm{RO}}^4 \, \underline{\underline{I}}\, \underline{\underline{K}}_{\parallel}^2 \\ \underline{\underline{I}} & \underline{\underline{\Omega}}(\omega) \end{bmatrix} \begin{bmatrix} \dfrac{\underline{v}_z(\mathbf{k},\omega)}{D} \\ \dfrac{}{\nabla c}\,\underline{c}(\mathbf{k},\omega) \end{bmatrix} = \begin{bmatrix} \underline{F}_1(\mathbf{k},\omega) \\ \underline{F}_2(\mathbf{k},\omega) \end{bmatrix}, \tag{S41}$$

where the diagonal matrices of momenta are

$$\underline{\underline{K}}^4 = \mathrm{diag}\left(k_n^4\right), \qquad \underline{\underline{K}}^2 = \mathrm{diag}\left(k_n^2\right), \qquad \underline{\underline{K}}_{\parallel}^2 = \mathrm{diag}\left(k_{n\parallel}^2\right), \tag{S42}$$

with $k_n^2$ and $k_{n\parallel}^2$ given by Eqs. (S22) and (S23), respectively, and

$$\underline{\underline{\Omega}}(\omega) = \mathrm{diag}\left(k_n^2 + \frac{i\omega}{D}\right) = \underline{\underline{K}}^2 + \frac{i\omega}{D}\underline{\underline{1}}, \tag{S43}$$

where $\underline{\underline{1}}$ denotes the identity matrix in mode space. The vectors of fluctuation amplitudes are

$$\underline{v}_z(\mathbf{k},\omega) = \left[\delta v_z^{(0)}(\mathbf{k},\omega), \delta v_z^{(1)}(\mathbf{k},\omega), \delta v_z^{(2)}(\mathbf{k},\omega), \ldots\right]^{\top}, \tag{S44}$$

$$\underline{c}(\mathbf{k},\omega) = \left[\delta c^{(0)}(\mathbf{k},\omega), \delta c^{(1)}(\mathbf{k},\omega), \delta c^{(2)}(\mathbf{k},\omega), \ldots\right]^{\top}. \tag{S45}$$

Crucially, $\underline{\underline{I}}$ is the matrix of cross-integrals between the sine and cosine eigenfunctions,

$$[\underline{\underline{I}}]_{nm} = \frac{2}{L}\int_0^L \mathrm{d}x\, \sin\left(\frac{n\pi x}{L}\right)\cos\left(\frac{m\pi x}{L}\right) = \begin{cases} 0 & \text{for} \quad n = m, \\ \dfrac{2n}{\pi}\,\dfrac{1-(-1)^{n+m}}{n^2-m^2} & \text{for} \quad n \neq m. \end{cases} \tag{S46}$$

Since $\underline{\underline{I}}$ has non-vanishing off-diagonal terms, different modes are coupled to each other. This precludes deriving the finite-size contribution using the Poisson resummation and Mellin-transform methods described in Sec. S.V and Sec. S.VI, respectively.

## S.V. NUMERICAL CALCULATION OF THE SCALING FUNCTION $\Psi(X)$

In this section, we describe the application of the Poisson resummation formula to the numerical evaluation of the finite-size contribution to the scaling function. We recall that this resummation formula [8] states that, for a generic smooth function $f$,

$$\sum_{n\in\mathbb{Z}} f(n) = \sum_{p\in\mathbb{Z}} \int_{\mathbb{R}} \mathrm{d}u\, e^{-2\pi i p u} f(u). \tag{S47}$$

For an even function $f(u) = f(-u)$, this expression can be simplified and takes the form

$$\sum_{n=1}^{\infty} f(n) = \frac{1}{2}\left[\sum_{n\in\mathbb{Z}} f(n) - f(0)\right] = \frac{1}{2}\int_{\mathbb{R}} \mathrm{d}u\, f(u) - \frac{1}{2}f(0) + \sum_{p=1}^{\infty}\int_{\mathbb{R}} \mathrm{d}u\, \cos(2\pi p u) f(u). \tag{S48}$$

This formula can be applied to the expression for the scaling function in Eqs. (17) and (18) in the main text by considering

$$f(u) = \frac{1}{X}\int \frac{\mathrm{d}q_y \mathrm{d}q_z}{(2\pi)^2}\left\{1+\frac{\left[(u\pi/X)^2+q_y^2+q_z^2\right]^3}{(u\pi/X)^2+q_y^2}\right\}^{-1}. \tag{S49}$$

After denoting $q_x = (u\pi/X)$, we obtain

$$\Psi(X) = \frac{1}{8\sqrt{2\pi}}\frac{\Gamma(7/4)}{\Gamma(9/4)} - \frac{1}{8\pi X} + 2\sum_{p=1}^{\infty}\int \frac{\mathrm{d}^3\mathbf{q}}{(2\pi)^3}\frac{\cos(2pXq_x)}{1+q^6/(q^2-q_z^2)}. \tag{S50}$$

Noticing that the first term in this expression corresponds to the bulk contribution, the finite-size contribution turns out to be

$$\Psi^{(\mathrm{fs})}(X) = -\frac{1}{8\pi X} + 2\sum_{p=1}^{\infty}\int \frac{\mathrm{d}^3\mathbf{q}}{(2\pi)^3}\frac{\cos(2pXq_x)}{1+q^6/(q^2-q_z^2)}. \tag{S51}$$

The second term in the expression above can be cast in the form

$$2\sum_{p=1}^{\infty}\int \frac{\mathrm{d}^3\mathbf{q}}{(2\pi)^3}\frac{\cos(2pXq_x)}{1+q^6/(q^2-q_z^2)} = \frac{1}{4\pi^2}\sum_{p=1}^{\infty}\int_0^{2\pi}\mathrm{d}\varphi\int_0^1 \mathrm{d}u\,\alpha^3(u,\varphi)e^{-\sqrt{2}pXu\alpha(u,\varphi)}\sin\left(\frac{\pi}{4}-\sqrt{2}pXu\alpha(u,\varphi)\right), \tag{S52}$$

where

$$\alpha(u,\varphi) = \left(\cos^2\varphi + u^2\sin^2\varphi\right)^{1/4}. \tag{S53}$$

We note that each term in the $p$-series in Eq. (S52) is expressed in terms of a double integral over a bounded subset on $\mathbb{R}^2$. Hence, it can be evaluated independently by straightforward numerical integration. The sum of the resulting values up to some $p_{\max} > 1$ provides an approximation of the $p$-series which is increasingly accurate as $p_{\max}$ increases:

$$2\sum_{p=1}^{\infty}\int \frac{\mathrm{d}^3\mathbf{q}}{(2\pi)^3}\frac{\cos(2pXq_x)}{1+q^6/(q^2-q_z^2)} \simeq 2\sum_{p=1}^{p_{\max}}\int \frac{\mathrm{d}^3\mathbf{q}}{(2\pi)^3}\frac{\cos(2pXq_x)}{1+q^6/(q^2-q_z^2)}. \tag{S54}$$

Importantly, the summand is a decreasing function of $pX$. Hence, one can expect the truncated sum to approximate the series well for $X \gg 1/p_{\max}$. On the other hand, the values obtained within this approximate scheme may deviate from the actual values of $\Psi^{(\mathrm{fs})}(X)$ as $X$ decreases. In Fig. 2 in the main text, we plotted $\Psi^{(\mathrm{fs})}$ obtained from the summation truncated at $p_{\max} = 150$. To independently evaluate the function for $X \ll 1$, we employed the asymptotic analysis based on the Mellin transform discussed in, c.f., Section S.VI. Crucially, the two results differ from each other solely within the interval $0 \leq X \lesssim 0.25$. This discrepancy is a consequence of the truncation of the sum and, in fact, the extension of this interval can be reduced by increasing $p_{\max}$.

## S.VI. ASYMPTOTIC BEHAVIORS OF $\Psi^{(\mathrm{fs})}(X)$

In order to determine analytically the asymptotic behaviors of $\Psi^{(\mathrm{fs})}$ for $X \gg 1$ and $X \ll 1$, we employ the Mellin transform, which, for a smooth function $f(u)$ defined for $u > 0$, is given by [9]

$$\hat{f}(s) = \int_0^{\infty}\mathrm{d}u\,f(u)\,u^{s-1}, \tag{S55}$$

where $s \in \mathbb{C}$ (not to be confused with the compressibility parameter $s$ defined in Eq. (26) in the main text). The inverse transform is provided by

$$f(u) = \int_{\mu-i\infty}^{\mu+i\infty}\frac{\mathrm{d}s}{2\pi i}\hat{f}(s)u^{-s}, \tag{S56}$$

where $\mu \in \mathbb{R}$ is chosen within the domain of analyticity of $\hat{f}(s)$. We note that

$$\sum_{n=1}^{\infty} f(n) = \sum_{n=1}^{\infty} \int_{\mu-i\infty}^{\mu+i\infty} \frac{\mathrm{d}s}{2\pi i} \hat{f}(s) n^{-s} = \int_{\mu-i\infty}^{\mu+i\infty} \frac{\mathrm{d}s}{2\pi i} \hat{f}(s)\zeta(s), \tag{S57}$$

where $\zeta(s) = \sum_{n=1}^{\infty} n^{-s}$ is the Riemann zeta function and $\mu$ in the last expression is chosen within the domain of analyticity of $\hat{f}(s)\zeta(s)$. We recall here, for later convenience, that $\zeta(s)$ has a simple pole with residue 1 for $s = 1$.

In the following two subsections, we first consider the case of interest in this work, in which the non-equilibrium density gradient is maintained parallel to the plates providing the spatial confinement. Then we show how to recover, from the present formalism, the case in which the gradient is perpendicular to the plates, which was previously investigated in the literature Refs. [10–12].

### A. Gradient parallel to the plates

In the case of the gradient parallel to the plates, the wavevectors satisfy Eqs. (S22) and (S23). We choose $f(u)$ as in Eq. (S49) and denoting again $q_x = (u\pi/X)$, we find

$$\hat{f}(s) = \frac{1}{4\pi^2 X} \left(\frac{X}{\pi}\right)^s \int \mathrm{d}^3\mathbf{q}\, \vartheta(q_x) \frac{q_x^{s-1}}{1 + q^6/(q^2 - q_z^2)}, \tag{S58}$$

where $\vartheta(x)$ is the Heaviside step function. We evaluate the integral above in spherical coordinates

$$\begin{aligned} q_x &= q \sin\theta \sin\phi, \\ q_y &= q \sin\theta \cos\phi, \\ q_z &= q \cos\theta. \end{aligned} \tag{S59}$$

Here, in order to guarantee that $q_x > 0$, we consider $\phi \in [0, \pi]$ and $\theta \in [0, \pi]$. Solving the corresponding integrals leads to

$$\hat{f}(s) = \frac{1}{16X} \left(\frac{X}{\pi}\right)^s \frac{\Gamma\left(\frac{s}{2}\right)\Gamma\left(\frac{3s+4}{4}\right)}{\Gamma\left(\frac{s+1}{2}\right)\Gamma\left(\frac{3s+6}{4}\right)\cos\left(\frac{\pi s}{4}\right)} \tag{S60}$$

Utilizing Eq. (S57), the above expression renders

$$\Psi(X) = \frac{1}{16X} \int_{\mu-i\infty}^{\mu+i\infty} \frac{\mathrm{d}s}{2\pi i} \left(\frac{X}{\pi}\right)^s \frac{\zeta(s)\Gamma\left(\frac{s}{2}\right)\Gamma\left(\frac{3s+4}{4}\right)}{\Gamma\left(\frac{s+1}{2}\right)\Gamma\left(\frac{3s+6}{4}\right)\cos\left(\frac{\pi s}{4}\right)}, \tag{S61}$$

where $\mu \in ]1, 2[$. Evaluating the poles of the expression above, we find the asymptotic behavior of the scaling function reported in Eqs. (22) and (23) in the main text for $X \gg 1$ and $X \ll 1$, respectively.

### B. Gradient perpendicular to the plates

If the concentration gradient is perpendicular to the plates and parallel to the gravitational acceleration, i.e., if we consider the confinement in the $z$-direction with $z \in [0, L]$ in Eqs. (3) and (4) in the main text, the wavevector of the fluctuations is quantized along the $z$-direction according to

$$k_n^2 = k_x^2 + k_y^2 + \left(\frac{n\pi}{L}\right)^2 \quad \text{and} \quad k_\parallel^2 = k_x^2 + k_y^2. \tag{S62}$$

In this geometry, the scaling function $\Psi_\perp(X)$ corresponding to $\Psi(X)$ is given by

$$\Psi_\perp(X) = \frac{1}{k_{\mathrm{RO}}^3} \frac{1}{L} \sum_{n=1}^{\infty} \int \frac{\mathrm{d}^2\mathbf{k}}{(2\pi)^2} \left[1 + \frac{k_n^6}{k_{\mathrm{RO}}^4 k_\parallel^2}\right]^{-1} = \frac{1}{X} \sum_{n=1}^{\infty} \int \frac{\mathrm{d}^2\mathbf{q}}{(2\pi)^2} \frac{1}{1 + q_n^6(X)/q_\parallel^2}, \tag{S63}$$

where $\mathrm{d}^2\mathbf{k} = \mathrm{d}k_x \mathrm{d}k_y$ and $\mathrm{d}^2\mathbf{q} = \mathrm{d}q_x \mathrm{d}q_y$. In the second equation, we rescaled the wavevectors according to $\mathbf{k} \mapsto \mathbf{q} = \mathbf{k}/k_{\mathrm{RO}}$, which yields

$$q_n^2(X) = q_x^2 + q_y^2 + \left(\frac{n\pi}{X}\right)^2 \quad \text{and} \quad q_\parallel^2 = q_x^2 + q_y^2. \tag{S64}$$

Then, in order to investigate the finite-size contribution $\Psi_\perp^{(\text{fs})}(X)$ to the scaling function, i.e.,

$$\Psi_\perp^{(\text{fs})}(X) = \Psi_\perp(X) - \lim_{X\to\infty} \Psi_\perp(X), \tag{S65}$$

we utilize the Mellin transform given by Eqs. (S55) and (S56). In particular, we set

$$f(u) = \frac{1}{X}\int \frac{\mathrm{d}^2\mathbf{q}}{(2\pi)^2}\left\{1 + \frac{\left[q_x^2+q_y^2+(u\pi/X)^2\right]^3}{q_x^2+q_y^2}\right\}^{-1}; \tag{S66}$$

after appropriate rescaling, its Mellin transform takes the form

$$\hat{f}(s) = \frac{1}{4\pi^2 X}\left(\frac{X}{\pi}\right)^s \int \mathrm{d}^3\mathbf{q}\,\vartheta(q_z)\frac{q_z^{s-1}}{1+q^6/(q^2-q_z^2)}, \tag{S67}$$

where we denoted $q_z = u\pi/X$. Again, we solve the integral in the spherical coordinates defined in Eq. (S59). However, in order to guarantee $q_z > 0$, we consider $\phi \in [0, 2\pi]$ and $\theta \in [0, \pi/2]$. Evaluating the resulting integral renders the Mellin transform

$$\hat{f}(s) = \frac{1}{16X}\left(\frac{X}{\pi}\right)^s \frac{\Gamma\left(\frac{s}{2}\right)\Gamma\left(\frac{s+6}{4}\right)}{\Gamma\left(\frac{3s+6}{4}\right)\cos\left(\frac{\pi s}{4}\right)}. \tag{S68}$$

This, using the observation in Eq. (S57), leads to the expression for the scaling function $\Psi_\perp$ in Eq. (S63) of the form

$$\Psi_\perp(X) = \frac{1}{16X}\int_{\mu-i\infty}^{\mu+i\infty}\frac{\mathrm{d}s}{2\pi i}\left(\frac{X}{\pi}\right)^s \frac{\zeta(s)\Gamma\left(\frac{s}{2}\right)\Gamma\left(\frac{s+6}{4}\right)}{\Gamma\left(\frac{3s+6}{4}\right)\cos\left(\frac{\pi s}{4}\right)}. \tag{S69}$$

As in the case of the previous subsection, the asymptotic behavior of this expression can be easily determined by evaluating its poles, which leads to

$$\Psi_\perp(X) = \begin{cases} \dfrac{1}{8\sqrt{2\pi}}\dfrac{\Gamma(7/4)}{\Gamma(9/4)} - \dfrac{1}{16X} + \mathcal{O}(X^{-7}) & \text{for} \quad X \gg 1, \\ \dfrac{X}{48\pi} + \mathcal{O}(X^5) & \text{for} \quad X \ll 1. \end{cases} \tag{S70}$$

We note that the $X$-independent term in $\Psi_\perp(X)$ for $X \gg 1$ renders the bulk contribution obtained in the Poisson-resummation calculation in Eq. (S50), which was derived in the geometry where the concentration gradient is parallel to the confining wall. This suggests that, as expected, the bulk contribution is independent of the geometry under consideration. Being a bulk term, it is also expected to be insensitive to the boundary conditions imposed on the fluctuations. A systematic verification of this latter expectation is left for future work. Note that the coefficient of the linear growth of $\Psi_\perp(X)$ for $X \ll 1$ in Eq. (S70) is

$$\lim_{X\to 0}\frac{\Psi_\perp(X)}{X} = \frac{1}{48\pi}. \tag{S71}$$

Recall that the limit $X \to 0$ corresponds to considering the case $g \to 0$ in which the gravity is absent. The above expression reproduces the results obtained in Refs. [10–12]. This work investigated the non-equilibrium pressure resulting from the confinement of temperature fluctuations with Dirichlet BCs in the considered geometry in the absence of gravity with free BCs imposed on the velocity field fluctuations. We note that the analysis of the problem in the presence of temperature fluctuations and in the absence of gravity is mathematically very similar to the case in which isothermal concentration fluctuations are considered (still in the absence of gravity), with the only difference being in the specific thermodynamic and transport coefficients which eventually determine the overall amplitude of the effect [5]. This is the reason why the analysis presented here renders the results of the references mentioned above.

## S.VII. ESTIMATE OF THE LOGARITHMIC DERIVATIVE OF THE COMPRESSIBILITY FACTOR

In this section we provide an estimate of the value of the parameter $s$, defined in Eq. (26) of the main text and entering the expression for the Casimir pressure for a colloidal suspension in Eq. (25), based on the available data from

light-scattering experiments. Following Ref. [13] (see Eq. (2) therein), we note that, in equilibrium, the intensity $I$ of the scattered light is related to the static structure factor $S_E$ defined in Eq. (S7), by

$$I = A_0 \left(\frac{\partial m}{\partial c}\right)^2_{p,T} S_E, \tag{S72}$$

where $A_0$ is a proportionality constant, which depends on the parameters of the scattering experiment, $m$ is the refractive index of the suspension, and $c$ is the concentration of the solute in the solvent. Accordingly, focusing on the dependence of the intensity $I$ on the concentration $c$ at constant temperature and pressure, the previous equation can be written as

$$I(c) \propto \left(\frac{\partial m}{\partial c}\right)^2_{p,T} S_E \propto \left(\frac{\partial m}{\partial c}\right)^2_{p,T} \frac{\chi(c)}{\rho(c)}, \tag{S73}$$

where we neglect all $c$-independent proportionality constants. In the case of colloidal suspensions, we assume that the osmotic susceptibility $\chi$ takes the form reported in Eq. (S3), which can be inverted in order to express the compressibility factor $Z$ defined there in terms of the ratio $c/\chi$. In turn, $\chi$ is related to the experimentally measurable intensity $I$ via Eq. (S73) and thus we find that

$$Z(\phi(c)) = B \left(\frac{\partial m}{\partial c}\right)^2_{p,T} \frac{c}{\rho(c)I(c)}, \tag{S74}$$

where $B$ is an overall multiplicative constant (encompassing $m_p$ and the optical prefactor $A_0$) that is not independently calibrated in the experiment. Since $B$ does not depend on the concentration $c$, it contributes to the dependence of $\log Z$ on $\log\phi$ only via an inconsequential additive constant, which does not affect the derivative $s$ of the former with respect to the latter (see Eq. (26) in the main text). Accordingly, $s$ can be determined from the logarithmic derivative of $(\partial m/\partial c)^2_{p.T}\, c/[\rho(c)I(c)]$ with respect to $\phi$. In Fig. S1(a), we present the experimental data for the intensity $I$ as a function of $c$, as obtained in Ref. [13] and reported in Fig. 3 of its electronic supplemental information. These data correspond to reference differential dynamic microscopy measurements performed at uniform equilibrium concentration (i.e., in the absence of an imposed concentration gradient), ensuring that the measured intensity reflects the static structure factor $S_E$ in the absence of non-equilibrium concentration fluctuations.

We can now relate the volume fraction $\phi$ of the solute to its mass fraction $c$ according to

$$c = \frac{\rho_p\phi}{\rho_p\phi + \rho_s(1-\phi)} \quad \text{or, conversely,} \quad \phi = \frac{\rho_s c}{\rho_s c + \rho_p(1-c)}, \tag{S75}$$

where the material density of silica particles is $\rho_p \simeq 2.2\,\mathrm{g/cm^3}$, and $\rho_s$ is the material density of the solvent, i.e., in the considered case, water, with $\rho_s \simeq 1\,\mathrm{g/cm^3}$. Similarly, the density $\rho(c)$ of the suspension can be expressed in terms of the material densities and the mass fraction according to

$$\rho(c) = \frac{\rho_p\rho_s}{\rho_s c + \rho_p(1-c)}. \tag{S76}$$

In order to estimate $(\partial m/\partial c)_{p,T}$, we invoke the Lorentz-Lorenz relation [14] which, for a binary mixture characterized by the volume fraction $\phi$, takes the form [15]

$$\frac{m^2(\phi)-1}{m^2(\phi)+2} = R(\phi) \equiv R_s + \phi(R_p - R_s), \tag{S77}$$

where $R_s$ and $R_p$ are the molar refractivities of the pure components, i.e., water and silica, respectively. From the previous equation, one easily finds

$$m(\phi) = \sqrt{\frac{1+2R(\phi)}{1-R(\phi)}}. \tag{S78}$$

We note that the refractive index of pure water and of silica are $m_s \simeq 1.33$ and $m_p \simeq 1.45$, respectively, which correspond to refractivities $R_s \simeq 0.204$ and $R_p \simeq 0.268$. Equation (S78) allows us to calculate

$$\left(\frac{\partial m}{\partial c}\right)_{p,T} = \left(\frac{\partial \phi}{\partial c}\right)_{p,T} \left(\frac{\partial m}{\partial \phi}\right)_{p,T} = \frac{3(R_p - R_s)}{2m(\phi)[1-R(\phi)]^2} \left(\frac{\partial \phi}{\partial c}\right)_{p,T}, \tag{S79}$$

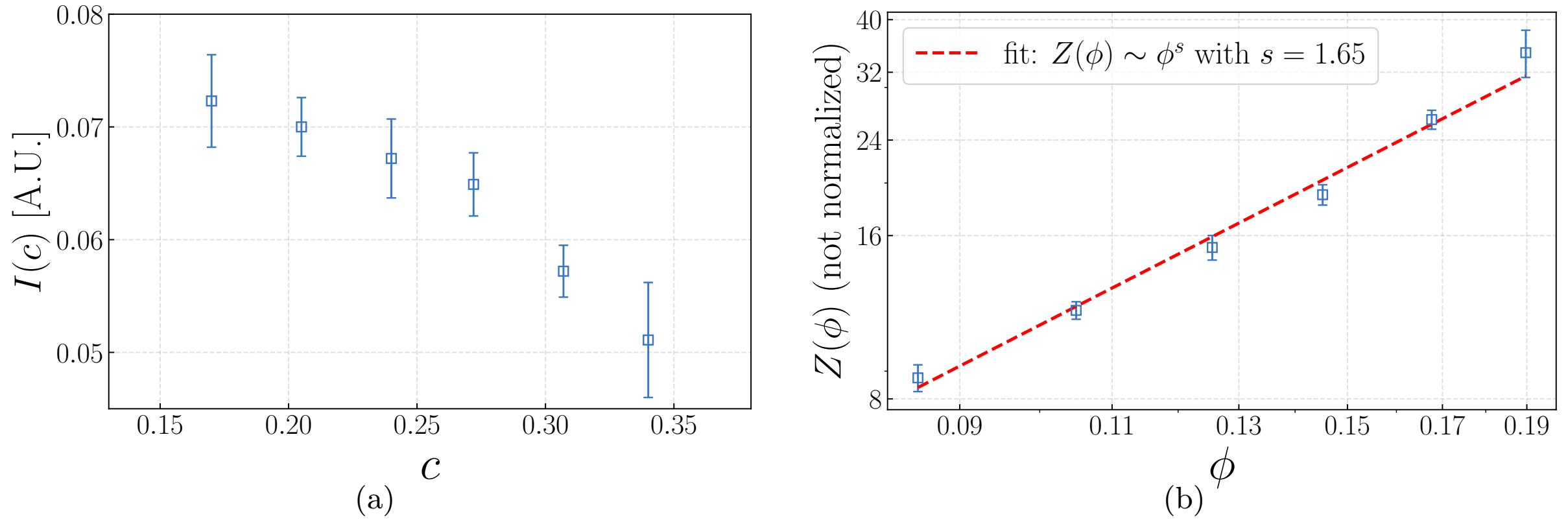


FIG. S1. (a) Scattered intensity $I(c)$ as function of the mass fraction $c$ of the silica particles in the colloidal dispersion, obtained in differential dynamic microscopy measurements performed at uniform equilibrium concentration. The data were digitized from Fig. 3 of the electronic supplemental information of Ref [13]. (b) Compressibility factor $Z$ as a function of the volume fraction $\phi$, plotted on a log-log scale, inferred from the experimentals data of panel (a), as explained in the text. The red dashed line corresponds to a linear fit, the slope of which renders the logarithmic derivative of $Z$ with respect to $\phi$, i.e., the parameter $s = 1.65 \pm 0.1$ which enters the expression of the Casimir pressure in colloidal suspensions, reported in Eq. (25) in the main text.

where $(\partial m/\partial c)_{p,T}$ on the r.h.s. is given by the derivative of Eq. (S75), i.e.,

$$\left(\frac{\partial \phi}{\partial c}\right)_{p,T} = \frac{[\rho_p \phi + \rho_s (1-\phi)]^2}{\rho_p \rho_s}. \tag{S80}$$

The relations in Eqs. (S75), (S76), (S79) and (S80) allow us to determine $Z$ (up to the proportionality constant $B$) as a function of $\phi$, based on the data reported in Fig. S1(a). The log-log plot of the resulting $Z(\phi)$ as a function of $\phi$ is presented in Fig. S1(b), together with a dashed red straight line, the slope of which corresponds to the parameter $s$, defined in Eq. (26) of the main text. From the fit to the experimental data, we obtain $s = 1.65 \pm 0.10$.




\end{document}